\documentclass[prd,aps,amsfonts,showpacs,longbibliography,notitlepage,twocolumn,superscriptaddress,nofootinbib]{revtex4-1}
\usepackage[dvipsnames]{xcolor}
\usepackage{multirow}
\usepackage{tabularx}
\usepackage{amsmath}
\usepackage{graphicx}

\usepackage[colorlinks=true, urlcolor=violet, linkcolor=blue, citecolor=red, hyperindex=true, linktocpage=true]{hyperref}

\newcolumntype{Y}{>{\centering\arraybackslash}X}
\newcolumntype{M}[1]{>{\centering\arraybackslash}m{#1}}

\newcommand{\andy}[1]{{\color{Rhodamine} Andy:~#1}}

\newcommand{\XYH}[1]{{\color{ForestGreen} XYH:~#1}}

\newcommand{\comment}[1]{}

\usepackage{comment}

\newcommand{\vect}[1]{\boldsymbol{#1}}
\newcommand{\im}{\mathrm{Im}~}
\newcommand{\re}{\mathrm{Re}~}

\newcommand{\cO}{\mathcal{O}}
\newcommand{\ud}{\mathrm{d}}
\newcommand{\p}{\partial}
\newcommand{\e}{\mathrm{e}}
\newcommand{\ii}{\mathrm{i}}

\begin{document}
\title{Disordered quantum critical fixed points from holography}

\author{Xiaoyang Huang}
\email{xiaoyang.huang@colorado.edu}
\affiliation{Department of Physics and Center for Theory of Quantum Matter, University of Colorado, Boulder, CO 80309, USA}

\author{Subir Sachdev}
\affiliation{Department of Physics, Harvard University, Cambridge, MA 02138, USA}

\author{Andrew Lucas}
\email{andrew.j.lucas@colorado.edu}
\affiliation{Department of Physics and Center for Theory of Quantum Matter, University of Colorado, Boulder, CO 80309, USA}

\begin{abstract}
 Using holographic duality, we present an analytically controlled theory of quantum critical points without quasiparticles, at finite disorder and finite charge density.  These fixed points are obtained by perturbing a disorder-free quantum critical point with relevant disorder whose operator dimension is perturbatively close to Harris-marginal.  We analyze these fixed points both using field theoretic arguments, and by solving the bulk equations of motion in holography.
 We calculate the critical exponents of the IR theory, together with thermoelectric transport coefficients.   Our predictions for the critical exponents of the disordered fixed point are consistent with previous work, both in holographic and non-holograpic models. 
\end{abstract}

\date{\today}

\maketitle

\section{Introduction}
Quantum field theory has proven to be a powerful tool to study and classify quantum phases of matter \cite{sachdevbook}.  In real experiments, of course, there is always disorder; the Harris criterion \cite{harris} determines whether such disorder qualitatively changes the IR fixed point (whether it is relevant or irrelevant).  When disorder is Harris-relevant, it is challenging to understand the \emph{intrinsically disordered} IR fixed points that arise. Existing constructions in higher dimensions are often analyzed close to fixed points with quasiparticles, such as free theories or large-$N$ vector models \cite{cardy1982,wen,Aharony:2015aea,thomson,raghu, Narovlansky:2018muj, Aharony:2018mjm,maciejko,Goldman:2019xrt,Ma:2022hso}.  The problem is especially difficult in theories at finite charge density, and/or with a Fermi surface, where controlled field theories of strongly interacting non-Fermi liquids are difficult to construct \cite{sachdevbook}.

 This Letter presents a controlled calculation, wherein we perturb a UV quantum critical point by Harris-relevant disorder, and analytically deduce the properties (critical exponents and transport coefficients) of the resulting compressible IR fixed point.  Our construction relies on holographic duality \cite{Hartnoll:2009sz,hartnoll_2018_holographic}, which maps certain models of ``matrix large-$N$" strongly interacting quantum field theories to classical gravity in one higher dimension.  These models holographically describe maximally chaotic \cite{hartnoll_2018_holographic,Maldacena:2015waa} field theories, which do not have any (known) quasiparticles.  Through a careful non-perturbative analysis of the nonlinear gravitational equations, we determine the scaling exponents and transport coefficients of the emergent IR fixed point, at finite disorder and finite density.

\section{Main result}
Let us summarize the main physical conclusions of the calculations.  We consider theories perturbed by disorder which couples to scalar operator $\mathcal{O}$: \begin{equation}\label{eq:disorder perturbation}
    S = S_0 + \int \mathrm{d}t \mathrm{d}^dx \; h(\vect x)\mathcal{O}(\vect x,t). 
\end{equation}
with $S_0$ a disorder-free action describing a quantum critical point with dynamical critical exponent $z$ and hyperscaling violation $\theta$.   $h(\vect x)$ is zero-mean Gaussian disorder: \begin{equation}\label{eq:variance}
    \overline{h(\vect x)h(\vect y)} \approx D \delta(\vect x-\vect y).
\end{equation}
The Harris criterion \cite{harris} tells us that disorder is relevant when the operator dimension $[D]>0$.  If the operator dimension of $\mathcal{O}$ is $\Delta$, defined by $\langle \mathcal{O}(x,0)\mathcal{O}(0,0)\rangle \sim |x|^{-2\Delta}$, then \cite{Lucas_prd2014} \begin{equation}
    [D] = -2\Delta + d -\theta + 2z.
\end{equation}
It is useful to write \begin{equation}
    \Delta = \frac{d-\theta}{2}+z-\nu, \label{eq:nu}
\end{equation}
so that $\nu=0$ corresponds to Harris-marginal disorder, while $\nu>0$ implies Harris-relevant disorder. For convenience, we also require $\cO$ not to be described by  alternate quantization in holography, so $\Delta>(d+z)/2$ \cite{Lucas_prd2014}.

We first discuss a minimal theory: a charge-neutral conformal field theory (CFT) in $d=1$ spatial dimension, perturbed by disorder as in (\ref{eq:disorder perturbation}), with $\nu=0$.  After a series of works \cite{adams,Hartnoll:2014cua,Hartnoll:2015rza,Aharony:2018mjm,Ganesan:2020wzm,Ganesan:2021gun}, it was shown that disorder is marginally \emph{irrelevant}: the scale-dependent disorder strength is captured by a beta function
\begin{equation}
    \beta_D = \frac{\mathrm{d}D}{\mathrm{d}\log E}= \frac{|C_{\cO\cO T}|}{C_{TT}}D^2; \label{eq:betaD}
\end{equation}
 $C_{\cO\cO T},C_{TT}$ are operator product expansion coefficients within the CFT.  

This Letter concludes this search for a disordered fixed point without quasiparticles as follows.  Just as the Wilson-Fisher fixed point can be perturbatively accessed in $d=3-\epsilon$ spatial dimensions \cite{wilsonfisher}, with $\epsilon$ perturbatively small, if we turn on a perturbatively small $\nu$ in (\ref{eq:nu}), 
\begin{equation} \label{eq:betanu}
    \beta_D  = \frac{|C_{\cO\cO T}|}{C_{TT}}D^2 - 2\nu D.
\end{equation}
This flow equation has a stable fixed point as $E\rightarrow 0$ if $\nu>0$: the value of disorder at the critical point is finite and non-zero, and takes the universal value \begin{equation}
    D^* = \frac{2\nu C_{TT}}{|C_{\cO\cO T}|}.
\end{equation}
Invoking a universal relation \cite{Aharony:2018mjm} between $D^*$ and $z^*$, valid for perturbations away from a conformal field theory, we obtain dynamical critical exponent
\begin{equation}\label{eq:d=1 cft fixed point}
    z^* = 1 + \frac{|C_{\cO\cO T}|}{C_{TT}}D^* = 1+2\nu .
\end{equation}

The argument above can be justified both using our holographic models, and using conformal perturbation theory to derive the exact prefactor of (\ref{eq:betaD}): see Appendix \ref{app:field theory} for the latter.  However, we do not know any field theoretic tools to generalize (\ref{eq:d=1 cft fixed point}) to perturbations of scaling theories where $z\ne 1$.  Yet these $z\ne 1$ theories include many interesting models of strange metals \cite{sachdevbook}. In contrast, we can more naturally generalize this argument to holographic models of a quantum critical point in $d$ spatial dimensions, at finite density $\rho$ of a conserved U(1) charge.  We take the exponents $z> \max(1+\theta/d,\theta)$ and $\theta \le d-1$, so that the holographic model obeys bulk energy conditions \cite{hartnoll_2018_holographic}. We then add Harris-relevant disorder through \eqref{eq:disorder perturbation}, satisfying \eqref{eq:variance} and \eqref{eq:nu} with $1\gg \nu>0$.  The system flows to a disordered IR fixed point characterized by a new set of scaling exponents $z^*,\theta^*$:
\begin{align}\label{eq:fixed point main}
    z^*\approx z+\frac{2\nu}{d}(z-\theta),\quad \theta^*=\theta.
\end{align}
While the hyperscaling violation $\theta$ remains the same as that in the disorder-free critical point for any $\nu$, the dynamical exponent $z$ will increase linearly in $\nu$ at the leading order. 

We have calculated the ac electrical conductivity $\sigma(\omega)$  at finite density IR fixed points.  We find (schematically) that \begin{equation} \label{eq:sigmamain}
    \sigma(\omega) \sim  \frac{K T^{-\frac{2+d-\theta^*}{z^*}}}{1-\mathrm{i}\omega \tau}+ F\left({\omega}/{T}\right)\omega^{2+\frac{d-\theta^*-2}{z^*}},
\end{equation}
where $K\sim \rho^2/D^*$ is a temperature-independent constant, and $F$ is a scaling function.  When $z^* < 2+d-\theta^*$, we find that $\tau T$ scales anomalously (diverges) as $T\rightarrow 0$: see (\ref{eq:tauT}).   If $\omega \ll T$, therefore, there is a sharp Drude peak, and the first term in (\ref{eq:sigmamain}) dominates.  The physical reason for this Drude peak is that the IR fixed point has perturbatively weak disorder ($D^*\sim \nu$), so the low frequency conductivity will be dominated by slow momentum relaxation: this is called a ``coherent" contribution to transport \cite{Hartnoll:2014lpa}.  The lifetime of momentum $\tau$ can be calculated using established methods \cite{Lucas:2015vna}, and we argue that it can be sensitive to UV thermodynamic data.  Hence, although the \emph{static} properties of the IR fixed point are universal, the width of any Drude peak is not. If $z^* \ge 2+d-\theta^*$, $\tau \lesssim 1/T$ would naively be sub-Planckian, so our conclusion is that that there is no well-defined Drude peak: the frequency dependence of the second term in (\ref{eq:sigmamain}) is more important.   When $\omega \gg T$, the second term in (\ref{eq:sigmamain}) dominates.  This is called the ``incoherent" conductivity, and is associated with current-relaxing dynamics decoupled from momentum relaxation.  The incoherent conductivity of the IR fixed point theory is universal and exhibits Planckian $\omega/T$ scaling; the function $F$ is insensitive to UV physics.


\section{Holography}
Having summarized the physics of the disordered fixed points, let us explain the holographic models we studied. In general, holography (``AdS/CMT") \cite{hartnoll_2018_holographic} is a powerful framework for building toy models of quantum matter without quasiparticles by mapping the physics on to a gravitational theory in one higher dimension.  Fields in the higher-dimensional ``bulk" theory correspond to low-dimension operators in the quantum field theory (QFT).  All QFTs have a stress tensor,  which is dual to the spacetime metric $g_{ab}$ in the bulk.  A finite density system requires a conserved U(1) current, dual to a bulk gauge field $A_a$.  A scalar field (dilaton) $\Phi$ in the bulk represents a scalar (spin-0) operators in QFT.   Following \cite{Lucas_prd2014,Huijse:2011ef,Lucas:2014sba}, we consider specifically the Einstein-Maxwell-Dilaton action in $d+2$-dimensional spacetime 
\begin{align}\label{eq:bulkS}
    S_0=\int \ud^{d+2} x \sqrt{-g}\left[\left(R-2(\p \Phi)^2-V(\Phi)\right)-\frac{Z(\Phi)}{4}F^2\right],
\end{align}
with coordinates $(r,t,\vect x)$. The bulk coordinate $r$ can intuitively be thought of as encoding energy scale in the QFT: the UV corresponds to $r\rightarrow 0$, while the IR is $r\rightarrow \infty$.  These EMD models are a standard holographic model capable of realizing fixed points for generic $z,\theta$.  To study Harris-relevant disorder, we introduce a bulk scalar field $\psi$, dual to the disorder operator $\cO$ in the QFT, and consider bulk action $S=S_0+S_\psi$, with
\begin{align}\label{eq:psi action}
    S_\psi = -\int \ud^{d+2} x \sqrt{-g}\left[ \frac{1}{2}(\p\psi)^2+\frac{B(\Phi)}{2} \psi^2  \right].
\end{align}
We emphasize that this differs from the usual strategy of studying disordered QFTs by introducing replicas \cite{cardy1982}: here, we study a single realization of the disorder, which is encoded by holographic duality in the \emph{boundary conditions}: $\psi(r\rightarrow 0,t,\vect x) \sim r^\# h(\vect x)$.  Note that the disordered boundary condition is random in $x$, but static in $t$.  

We will reveal the emergent IR fixed point by solving the nonlinear bulk equations of gravity, subject to these boundary conditions.  Details of the construction, including precise functional forms for $V,Z,B$, etc., are in Appendix \ref{app:holography}.  In the absence of disorder,
the metric is given by
\begin{align}\label{eq:emd metric}
    \ud s^2= \frac{1}{r^2}\left[ \frac{a(r)}{b(r)}\ud r^2-a(r)b(r)\ud t^2+\ud \vect{x}^2\right],
\end{align}
while the dilaton and gauge fields are
\begin{align}\label{eq:Phi A}
    \Phi = \Phi(r),\quad A = p(r)\ud t.
\end{align}
The scaling exponents $z,\theta$ are captured by the constants $a_0$ and $b_0$ in $a(r) \sim r^{a_0}$ and $b(r)\sim r^{b_0}$.  To study a finite density black hole, we can identify the charge density with
\begin{equation}
    \rho =  \left.-\frac{Z p'}{a r^{d-2}}\right. .
\end{equation}  

In the presence of spatially inhomogeneous $\psi$, an analytical solution of the classical bulk equations cannot be found.  Indeed, with hindsight, (\ref{eq:fixed point main}) shows that $a_0$ and $b_0$ will get linear corrections in $\nu$, which are non-perturbative corrections in $\nu$ to the actual bulk fields.  To understand how to solve these complicated bulk equations, let us begin with a physical picture for the radial evolution of the geometry from UV ($r=0$) to IR ($r=\infty$).  If the disorder is self-averaging (the geometry is, at leading order, independent of disorder realization), then the geometry must be approximately homogeneous in $x$: after averaging over disorder realizations, translation invariance is restored. The bulk geometry is constructed holographically by varying the action (\ref{eq:bulkS}) and solving the equation of motion for each field; e.g. for the metric, we obtain: \begin{align} \label{eq:einstein}
    R_{ab}  - \frac{R}{2}g_{ab} = \frac{1}{2}\left(T^A_{ab}+T^\Phi_{ab} + \overline{T^\psi_{ab}} \right),
\end{align}
where $T^{A,\Phi,\psi}_{ab}$ denote the \emph{bulk} stress tensors associated with each of these fields, and $\overline{\cdots}$ denotes disorder averaging.   We then solve for the bulk fields $a,b,p,\Phi$ non-perturbatively, assuming that they are sourced by the homogeneous $\overline{T^\psi_{ab}}$.  We make the general ansatz
\begin{subequations}\label{eq:ansatz}
\begin{align}
    a(r) &\approx \alpha_0 r^{a_0-\gamma_a(r)} ,\\
    b(r) &\approx \beta_0 r^{b_0-\gamma_b(r)} ,\\
    \Phi(r)& \approx c_\Phi(r) \log r,\\
    p(r)&\approx \pi_0 r^{p_0-\gamma_p(r)}.
\end{align}
\end{subequations}
which readily suggests a physical interpretation: $\gamma_{a,b,p}$ will encode the flow of critical exponents from the UV to IR fixed points.  

Plugging in (\ref{eq:ansatz}) into the homogenized bulk equation of motions, we obtain equations to solve for $\gamma_{a,b,p}$ and $c_\Phi$.  Together with the equation of motion for each Fourier mode $\psi(r,\boldsymbol{k})$, we can then solve for all bulk fields and obtain a self-consistent solution to (\ref{eq:einstein}).  While we leave most details of this calculation to Appendix \ref{app:holography}, let us describe the critical part of the calculation.  The bulk equations of motion imply that $c_\Phi$ remains constant and $\gamma_a\approx \gamma_b \approx \gamma_p= \gamma$, which in turn obeys \begin{align} \label{eq:gammaeq}
  \gamma &+  r\log r \gamma^\prime - AD r^{\frac{2d\nu}{d-\theta}-\frac{d}{z-\theta}\gamma } \notag \\
  &= \frac{(d-\theta)r}{d(d+z-\theta)}\partial_r  \left( \gamma +  r\log r \gamma^\prime\right).
\end{align}
 $A$ is a constant depending on $z$ and $\theta$.  Applying dominant balance to (\ref{eq:gammaeq}), the right hand side is negligible, and
\begin{align}\label{eq:gamma rel max full}
    \gamma(r) \approx \frac{z-\theta}{d\log r }\log\left[1+ AD \frac{(d-\theta)}{2\nu(z-\theta)} r^{\frac{2d\nu}{d-\theta}} \right].
\end{align}
The bulk geometry locally looks like a scaling geometry, with $z$ varying extremely slowly; this enables us to analytically solve for the eventual fixed point. Numerical solutions confirm that this fixed point is the only one consistent with an approximately homogeneous bulk geometry (see Appendix \ref{app:numerics}).

To illustrate what (\ref{eq:gamma rel max full}) implies, we define a \emph{dimensionless} effective disorder strength \begin{align}\label{eq:Dr flow}
    D_{\mathrm{eff}}\equiv D  r^{\frac{2d\nu}{d-\theta}- \frac{d}{z-\theta}\gamma} = \frac{Dr^{\frac{2d\nu}{d-\theta}}}{1+DA \frac{(d-\theta)}{2\nu(z-\theta)} r^{\frac{2d\nu}{d-\theta}}}.
\end{align}
Notice that $D_{\mathrm{eff}}\rightarrow 0$ as $r\rightarrow 0$, since disorder is Harris-relevant.  In the IR, \begin{equation}\label{eq:Dstar}
    D_{\mathrm{eff}} \rightarrow D^* = \frac{2\nu (z-\theta)}{A(d-\theta)}
\end{equation}
approaches a universal constant.  This is the disorder strength of \emph{exactly Harris marginal disorder} that supports the IR fixed point!  Since (\ref{eq:gamma rel max full}) implies that $\gamma = AD^*$ at the IR fixed point, we can solve for the IR critical exponents $z^*,\theta^*$, and we find (\ref{eq:fixed point main}). The crossover energy scale $E_c$ between the UV and IR fixed points occurs at the non-perturbatively large scale 
\begin{align}\label{eq:ec}
    E_c\sim \left(\frac{D}{\nu}\right)^{\frac{z^*}{2 \nu}},
\end{align}
emphasizing the non-perturbative nature of our (approximate) solution to the nonlinear bulk equations.  It is interesting that such a detailed analysis of the bulk equations is needed to reproduce what, in a field theoretic language (\ref{eq:betanu}), is a perturbative one-loop effect.

It remains to explain why the geometry is self-averaging \cite{Ganesan:2020wzm}.  While  at $O(D)$ the disorder contributed to a homogeneous source $\overline{T^\psi_{ab}}$ for gravity in (\ref{eq:einstein}), there will also be inhomogeneous source terms proportional to $h(\boldsymbol{k})h(\boldsymbol{q})$ with $\boldsymbol{k}+\boldsymbol{q}\ne \mathbf{0}$.  These inhomogeneous source terms would not matter if the left hand side of (\ref{eq:einstein}) was linear; since it is nonlinear in $g_{ab}$, such source terms do feed back and correct the metric beyond our ansatz.  However, to correct the disorder averaged metric, we will need at least two such powers of the source term, meaning that there are four factors of $h$.  Thus, the corrections to our approximation are $O(D^2)=O(\nu^2)$.  Since at the IR fixed point, disorder remains perturbatively small, this correction can be neglected at leading non-trivial order, thus justifying that the geometry is self-averaging at the perturbatively accessible fixed point.

We studied a charge-neutral critical point with a non-trivial hyperscaling violation $\theta\ne 0$. This is done by turning off the bulk gauge field ($A_a=0$); Lorentz invariance in the boundary directions demands $z=1$. The dilaton field will get renormalized ($c_\Phi$ is no longer constant), and the disordered IR fixed point has critical exponents (Appendix \ref{app:charge neutral})
\begin{subequations}\label{eq:FPrel no maxwell}
\begin{align}
    z^* &= 1+ \frac{6 \nu (1-\theta) (d-\theta)}{d (3 d+(\theta-5) \theta)},\\
    \theta^* &=\theta+\frac{2 \nu (\theta-1)  (d-\theta)}{d (3 d+(\theta-5) \theta)}\theta.
\end{align}
\end{subequations}
We see that $\theta$ is renormalized.
Interestingly, as long as $\theta\neq 0$, we have a different fixed point from \eqref{eq:fixed point main} by taking $z\to 1$ there, and this is because when $z\neq 1$, $\theta$ is not renormalized. Nevertheless, \eqref{eq:fixed point main} and \eqref{eq:FPrel no maxwell} agree in the CFT limit: $z=1$ and $\theta=0$.

Observe that \eqref{eq:fixed point main} and \eqref{eq:FPrel no maxwell} are consistent with the general expectation that disorder should become exactly marginal at the IR fixed point: if it was relevant, it would drive us to a new fixed point; if it was irrelevant, then the IR would not have finite disorder $D^*$! 
To confirm that the disorder is exactly Harris-marginal at the IR fixed point, we compute its scaling dimension $\Delta_{\mathrm{IR}}$.  In AdS space, the mass of a bulk field determines the dual operator's scaling dimension; for us, $\Delta_{\mathrm{IR}}$ is fixed by $B(\Phi)$.  Calculating $\Delta_{\mathrm{IR}}$ from $B(\Phi)$ and demanding that it is Harris-marginal ($\Delta_{\mathrm{IR}} = \frac{d-\theta^*}{2}+z^*$), we find the condition that 
\begin{align}\label{eq:IR marginal d}
    \frac{d}{z-\theta}(z^*-z) +\frac{2dz-d\theta}{(z-\theta)(d-\theta)}(\theta^* - \theta)=2\nu.
\end{align}
Obviously, \eqref{eq:fixed point main} and \eqref{eq:FPrel no maxwell} satisfy the above equation.

Previous literature \cite{Gubser-Rocha,Hartnoll:2012wm} has studied theories with $z/(-\theta)=\eta>0$ fixed, while $z\rightarrow\infty$.  Such theories are analyzed in Appendix \ref{app:local critical}.

\section{Conductivities}

We now discuss the thermoelectric transport properties of the disordered IR fixed point.  
We study the theory at temperatures $T\ll E_c$, whereby the geometry is approximately that of the IR fixed point, but contains a black hole horizon $r=r_+$ with Hawking temperature $T$.  This corresponds to modifying the geometry found in (\ref{eq:ansatz}) via \cite{hartnoll_2018_holographic} \begin{equation}
    b(r) \rightarrow b(r)\left(1-\left(\frac{r}{r_+}\right)^{d+\frac{dz^*}{d-\theta^*}}\right),
\end{equation}
where $T\sim r_+^{-\frac{dz^*}{d-\theta^*}}$.
At the horizon, the entropy density $s$ scales $s\sim r_+^{-d}\sim T^{\frac{d-\theta^*}{z^*}}$.

In general, if we apply a temperature gradient $-\nabla T \mathrm{e}^{-\mathrm{i}\omega t}$ and electric field $\mathbf{E} \mathrm{e}^{-\mathrm{i}\omega t}$, the charge current $\mathbf{J} \mathrm{e}^{-\mathrm{i}\omega t}$ and heat current $\mathbf{Q} \mathrm{e}^{-\mathrm{i}\omega t}$ are proportional to these sources: \begin{equation}
\left(\begin{array}{c} \mathbf{J} \\ \mathbf{Q} \end{array}\right) =   \left(\begin{array}{cc} \sigma(\omega) &\ \alpha(\omega) \\  T\alpha(\omega) &\ \bar\kappa(\omega)  \end{array}\right)  \left(\begin{array}{c} \mathbf{E} \\ -\nabla T \end{array}\right).
\end{equation}  
Let us first discuss the dc ($\omega=0$) conductivities. Via the membrane paradigm 
\cite{iqbal,Banks:2015wha}, we can evaluate them by analyzing the geometry at the horizon: see Appendix \ref{app:conductivity}. We find that the thermoelectric conductivities are all approximated by a Drude-like form, signifying that the transport coefficients are dominated by slow momentum relaxation: \cite{hartnoll_2018_holographic}    \begin{align}\label{eq:dc drude}
        \sigma_{\mathrm{dc}} \approx \frac{\rho^2}{\Gamma}, \;\;\; \alpha_{\mathrm{dc}} \approx \frac{\rho s}{\Gamma}, \;\;\; \bar\kappa_{\mathrm{dc}} \approx \frac{Ts^2}{\Gamma},
    \end{align}
where 
\begin{equation}\label{eq:Gamma}
    \Gamma \sim D^* T^{\frac{d-\theta^* + 2}{z^*}}.
\end{equation}
Remarkably, \eqref{eq:dc drude} agrees with the perturbative result in \cite{Lucas_prd2014} with Harris-marginal disorder (in the IR), again confirming the criterion in \eqref{eq:IR marginal d}.

Following \cite{Davison:2015bea,Blake:2015epa,Davison:2015taa}, we now analyze the subleading (in $D_*$) corrections to transport coefficients that describe transport decoupled from momentum relaxation.  As we show in Appendix \ref{app:conductivity}, in this holographic model such corrections to thermoelectric transport coefficients are captured by the open-circuit thermal conductivity \begin{align}
    \kappa_{\mathrm{dc}} &\equiv \bar\kappa_{\mathrm{dc}} - T \alpha_{\mathrm{dc}}^2\sigma_{\mathrm{dc}}^{-1}\sim T^{ \frac{z^*+d-\theta^*-2}{z^*}}.
\end{align}
In ordinary metals, one finds that $\kappa_{\mathrm{dc}}\sim T\sigma_{\mathrm{dc}}$ as $T\rightarrow 0$ with a precise prefactor (this is called the Wiedemann-Franz law) \cite{hartnoll_2018_holographic}; clearly, this is badly violated at these disordered fixed points, since
\begin{align}\label{eq:lorentz}
    \mathcal{L} \equiv \frac{\kappa_{\mathrm{dc}}}{T \sigma_{\mathrm{dc}}} \sim D^* T^{2\frac{d-\theta^*}{z^*}}
\end{align}
vanishes as $T\rightarrow 0$.  Anomalous scaling of $\mathcal{L}$ is not too surprising given that the leading order results (\ref{eq:dc drude}) exactly cancel in $\kappa_{\mathrm{dc}}$; indeed, it is the subleading corrections to $\sigma_{\mathrm{dc}}$ that are responsible for non-vanishing $\kappa_{\mathrm{dc}}$. One calls such contributions to thermoelectric transport ``incoherent" \cite{Hartnoll:2014lpa} as they are decoupled from slow momentum relaxation.

Let us now extend the discussion to ac ($\omega>0$) conductivity; for simplicity, we focus only on the electrical conductivity $\sigma(\omega)$.  Following \cite{Lucas:2015vna}, we find that there can be a Drude peak at low frequency $\omega \ll T$:  $\sigma(\omega) \sim \sigma_{\mathrm{dc}}/(1-\mathrm{i}\omega \tau)$, where $\tau = \mathcal{M}/\Gamma$.  We argue in Appendix \ref{app:conductivity} that $\mathcal{M}\sim T^0$ is a UV-sensitive quantity, implying that $\tau$ is \emph{not universal}, and exhibits anomalous temperature dependence: 
\begin{align} \label{eq:tauT}
    \tau \sim T^{-\frac{2+d-\theta^*}{z^*}}.
\end{align}
The holographic calculation of $\tau$ is only accurate if $\tau \gg 1/T$, so there is a sharp Drude peak only when $2+d-\theta^*\ge z^*$.
For theories that violate this inequality, we expect no sharp features in $\sigma(\omega)$ until the scale $\omega \sim T$.  For frequencies $\omega \gg T$, we find that the incoherent conductivity dominates the response function: \begin{equation}\label{eq:sigmahighfrequency}
    \sigma(\omega)\sim \omega^{2 + \frac{d-\theta^*-2}{z^*}}.
\end{equation}  

The various power laws found above are consistent with recent holographic scaling theories for IR fixed points at finite density  \cite{Davison:2015taa,Davison:2018nxm}.  Following \cite{Hartnoll:2015sea}, we assign the charge density operator an anomalous dimension $\Phi_\rho$:
\begin{align}
    [\rho] = d-\theta^*+\Phi_\rho.
\end{align}
Scaling analysis shows that $[\sigma_{\mathrm{dc}}] = d-\theta^* - 2+2\Phi_\rho$ \cite{Davison:2015taa}. In order to match with \eqref{eq:Gamma}, we find $\Phi_\rho=-d+\theta^*$, which implies $[\rho]=0$.  It has previously been observed \cite{Davison:2018nxm} that $[\rho]=0$ ensures the IR fixed point thermodynamics is consistent with scaling theories, and thus (\ref{eq:Gamma}) is consistent with this expectation.  A more careful analysis reveals that the incoherent conductivity has a \emph{different} IR scaling dimension: $[\sigma_{\mathrm{inc}}] = 3(d-\theta^*)-2+2z^*+2\Phi_\rho$ \cite{Davison:2018nxm}.  This is consistent with (\ref{eq:sigmahighfrequency}), and a direct calculation of the dc incoherent conductivity in  Appendix \ref{app:conductivity}.

\section{Outlook}
In this letter, we have analytically predicted the emergence of a disordered fixed point in a strongly interacting QFT, at either zero or finite density.  The exponents $z^*$ and $\theta^*$ are independent of UV disorder strength $D$, as are the dc thermoelectric transport coefficients.

The holographic formalism described here is versatile and could be used to study the emergence of finite disorder fixed points in more general settings, such as in background magnetic fields \cite{Hartnoll:2009sz}, or in the presence of non-trivial topological effects \cite{Landsteiner:2015lsa}. It would also be interesting to generalize to models with inhomogeneous charge disorder, where lattice constructions can reveal robust $T$-linear resistivity \cite{Balm:2022bju}.

We encourage further numerical work \cite{Dias:2015nua} to solve the fully inhomogeneous Einstein equations, and analyze the fixed points described here.  The most promising direction may be to focus on one-dimensional disordered systems; prior work \cite{Hartnoll:2015rza} constructed black holes with relevant disorder, but their value of $\nu = 3/4$ may be beyond the regime of validity of our perturbation theory.  At strong disorder, it may be possible for the horizon to fragment into disconnected pieces, a fascinating phenomenon whose implications for the boundary theory deserve further investigation \cite{Anninos:2013mfa,Horowitz:2014gva}.

Our result (\ref{eq:fixed point main}) may extend beyond holographic models.  In a (charge-neutral) large-$N$ vector model with non-disordered fixed point with $d=2$, $z=1$, $\theta=0$, the mass disorder at the critical point is relevant with $\nu = \frac{16}{3\pi^2 N}$; a recent calculation \cite{Goldman:2019xrt} found that $z^*\approx 1+\nu$ at the disordered fixed point. This agrees with (\ref{eq:fixed point main}).  It would be fascinating if our results can be extended to recent models \cite{Aldape22,Patel:2022gdh} of compressible, disordered non-Fermi liquids based on field theories, including those based on Sachdev-Ye-Kitaev models which display $\sigma \sim \omega^{-1}$.

\section*{Acknowledgements}
We acknowledge helpful comments from Blaise Gout\'eraux.
This work was supported by a Research Fellowship from the Alfred P. Sloan Foundation under Grant FG-2020-13795 (AL), by the Gordon and Betty Moore Foundation's EPiQS Initiative under Grant GBMF10279 (XYH, AL), and by the National Science Foundation under Grant No.~DMR-2245246 (SS).

\onecolumngrid

\newpage

\begin{appendix}
\section{Field theory perspective}\label{app:field theory}

Holographic duality describes some ``matrix large-$N$" theories using classical gravity \cite{hartnoll_2018_holographic}. What is important about this large-$N$ limit is that the theory is not quite described by a ``generalized free-field theory" where all non-trivial operator product expansion (OPE) coefficients are suppressed; see e.g. \cite{Aharony:2015aea}, where classical disorder remained exactly marginal at leading order in large $N$. Indeed, the simple description of the field theory is in terms of classical gravity in one higher dimension!

Nevertheless, with some minor adjustments, we can still use field theory ideas to understand the holographic results found in the main text.  For simplicity, we focus on the case $z=1$ and $\theta=0$, and assume charge neutrality, so that the more powerful technology of CFTs can be invoked. What follows is similar to \cite{Ganesan:2021gun}; however some technical steps differ.

The key observation is that given a bulk holographic action (we do not write down the counterterms at the boundary of the bulk spacetime for ease of presentation) \begin{align}
    S_{\mathrm{bulk}} = \mathcal{N}\int \mathrm{d}^{d+2}x \sqrt{-g} \left(R-2\Lambda + \frac{1}{2}(\partial \psi)^2 - \frac{1}{2}m^2\psi^2 - \alpha \psi^3 + \cdots \right), 
\end{align}
the OPE coefficients of the CFT are encoded in the prefactors of terms within $S_{\mathrm{bulk}}$.   In particular, if we expand the metric $g = g_{\mathrm{AdS}}+\delta g$ around an AdS background and $\psi = \delta \psi $ around 0, any OPE coefficient relating $n$ copies of the stress tensor $T$ and $m$ copies of the scalar operator $\mathcal{O}$ (dual to $\psi$) can be read off (schematically) as \begin{equation}
    C_{\mathcal{O} \mathcal{O}} \sim \frac{\delta^{2}S_{\mathrm{bulk}}}{ \delta \psi^2}, \;\;\;\;  C_{ T\mathcal{O} \mathcal{O}} \sim \frac{\delta^{3}S_{\mathrm{bulk}}}{\delta g \delta \psi^2}, \;\;\;\; \text{etc.}
\end{equation}
Note that here $\mathcal{N}$ scales as some power of $N$ depending on dimension (and can be derived from string theory \cite{} in some cases).  The upshot of this paragraph is that for a quadratic bulk action in (\ref{eq:bulkS}), the only leading order OPE coefficients at large $\mathcal{N}$ are $C_{TT}$, $C_{\mathcal{OO}}$, $C_{TTT}$, $C_{\mathcal{OO}T}$.


Now, let us try to understand a field theory where the listed OPE coefficients are the dominant ones, in the presence of quenched random-field disorder.  Unlike in holography, we now consider the replicated action
\begin{align}\label{eq:sn}
    S_n = \sum_A^n S_{0,A}  - \frac{D}{2} \sum_{AB} \int \ud^d x\ud t\ud t' ~ \cO_A(x,t) \cO_B(x,t').
\end{align}
This action arises upon using the replica trick to analyze (\ref{eq:disorder perturbation}), given disorder (\ref{eq:variance}).
For now let us assume marginal disorder $\Delta = d/2+1$ (i.e. $\nu=0$). Due to the double time integral, terms with $A=B$ at $t'\to t$ are singular. Now, perform the OPE at $t'\to t$, and we find following \cite{Aharony:2018mjm} that
\begin{align}
    \cO_A(x,t) \cO_B(x,t') \supset \frac{-|C_{\cO\cO T}|}{C_{TT}} \frac{1}{|t-t'|}T_{00,A}(x,t)\delta_{AB}+\cdots,
\end{align}
Plugging it into \eqref{eq:sn} and regularizing the time integral using a Wilsonian cutoff: $\Lambda^{-1} < |t-t^\prime| < b\Lambda^{-1}$ where $b\equiv \Lambda/E$, we find
\begin{align}
    \delta S_n = D \frac{|C_{\cO\cO T}|}{C_{TT}} \log b \sum_A \int \ud^d x\ud t ~T_{00,A}(x,t).
\end{align}
Observe that this correction to $\delta S_n$ appears as though we are simply rescaling the time coordinate: $t\to t Z_t$, where \cite{Aharony:2018mjm,Ganesan:2021gun}
\begin{align}\label{eq:zt}
    Z_t = 1+ D \frac{|C_{\cO\cO T}|}{C_{TT}} \log b.
\end{align}
Here comes the key observation: in holographic duality, as explained above, the ratio $C_{\mathcal{OO}T}/C_{TT}$ remains finite as $\mathcal{N}\rightarrow \infty$. To obtain the beta function $\beta_D$, we expand the action in perturbatively small $D$, and we find 
\begin{align}\label{eq:expand sn}
    \e^{-S_n} =& \e^{-\sum_A S_{0,A}}\Bigg[1+ \frac{D}{2} \sum_{AB} \int \ud^d x\ud t\ud t' ~\cO_A(x,t)\cO_B(x,t')+ (Z_t^{-2}-1)\frac{D}{2} \sum_{AB} \int \ud^d x\ud t\ud t' ~\cO_A(x,t)\cO_B(x,t')\nonumber\\
    & - \frac{D^2}{2} \frac{|C_{\cO\cO T}|}{C_{TT}} \log b \sum_{ABC} \int \ud^d x_1\ud t_1  \ud^d x_2\ud t_2\ud t_2' ~T_{00,A}(x_1,t_1)\cO_B(x_2,t_2)\cO_C(x_2,t_2')+O(D^3)\Bigg],
\end{align}
where the last term in the first line accounts for the rescaling of time in the double time integral.
Because the higher-order OPE coefficients such as $C_{\cO\cO\cO} = 0$ in the large-$N$ limit of interest, $\cO_A(x)\cO_B(0)\sim x^{-2\Delta}\delta_{AB}$, and
\begin{align}
    \sum_{AB} \cO_A\cO_B(x) \sum_{BC} \cO_C\cO_D(0)\sim \frac{4 n}{x^{2\Delta}}\sum_{AB} \cO_A\cO_B(0)
\end{align}
vanishes in the replica limit $n\to 0$. To see that $O(D^2)$ terms in \eqref{eq:expand sn} are the leading corrections that will renormalize $D$, we evaluate (part of) the second line following \cite{Aharony:2018mjm} as
\begin{align}
    \int \ud^d x_1 \ud t_1 \langle T_{00,A}(x_1,t_1)\cO_{A}(x_2,t_2)\cO_{A}(x_2,t_2') \rangle= -\frac{C_{\cO\cO}(d+2)}{|t_2-t_2'|^{d+2}} \sim -(d+2) \langle \cO_A(x_2,t)\cO_A(x_2,t') \rangle.
\end{align}
A natural RG scheme is to require that the one-point function of (\ref{eq:expand sn}) has no dependence on $\log b$: \begin{align}
    \langle 1  + \cdots \rangle = 1 + \frac{D+\delta D}{2}\sum_A \int \mathrm{d}^dx \mathrm{d}t\mathrm{d}t^\prime \langle \mathcal{O}_A(x,t)\mathcal{O}_A(x,t^\prime)\rangle.
\end{align}
Hence we can interpret the diffusion constant as flowing under RG, with
\begin{align}
    \delta D = -\frac{(d+2)|C_{\cO\cO T}|}{C_{TT}}D^2 \log b + \frac{2|C_{\cO\cO T}|}{C_{TT}}D^2 \log b = -\frac{d|C_{\cO\cO T}|}{C_{TT}}D^2 \log b.
\end{align}
Therefore, the beta-function reads
\begin{align}
    \beta_D =  -\frac{\p \delta D}{\p\log b}=  \frac{d|C_{\cO\cO T}|}{C_{TT}}D^2.
\end{align}
We see that the disorder is marginally irrelevant.

\section{Details of holographic models}\label{app:holography}

In the scaling limit, the expressions of \eqref{eq:emd metric} and \eqref{eq:Phi A}, or the bare value of \eqref{eq:ansatz},  are
\begin{subequations}\label{eq:scaling form}\begin{align}
    a(r) &= r^{-\frac{d(z-1)-\theta}{d-\theta}} ,\\
    b(r) &= r^{-\frac{d(z-1)+\theta}{d-\theta}} ,\\
    \Phi(r)& = \sqrt{\frac{d(d(z-1)-\theta)}{2(d-\theta)}} \log r,\\
    p(r)&= 2\frac{d(z-1)}{d-\theta} r^{-\frac{d(z+d-\theta)}{d-\theta}},
\end{align}\end{subequations}
and we take $\alpha_0=\beta_0=1$ for simplicity. For $\Phi$ to be real, we obtain the null energy condition 
\begin{align}\label{eq:null energy condition}
    \frac{d(z-1)-\theta}{d-\theta} \geq 0.
\end{align}
The potentials in the bulk action take the form $V(\Phi) = -V_0 \e^{-\beta\Phi}, Z(\Phi) = Z_0\e^{\alpha\Phi}$  with
\begin{subequations}\label{eq:bulkcoe}\begin{align}
    \alpha^2 &= \frac{8(d(d-\theta)+\theta)^2}{d(d-\theta)(d(z-1)-\theta)},\\
    \beta^2 &=\frac{8\theta^2}{d(d-\theta)(d(z-1)-\theta)},\\
    V_0 &= \frac{d^2(z+d-\theta)(z-1+d-\theta)}{(d-\theta)^2},\\
    Z_0&= \frac{(d-\theta)^2}{2d^2(z+d-\theta)(z-1)}.
\end{align}\end{subequations}
In order to maintain the scale invariance, the mass term for the scalar field has to take the form \cite{Lucas_prd2014}
\begin{align}\label{eq:mass term}
    B(\Phi) = B_0 \frac{b(r)}{a(r)} = B_0 \e^{-\beta \Phi}.
\end{align}
Since $b(r)/a(r)\sim r^{-2\theta/(d-\theta)}$, we see how the hyperscaling violation $\theta$ is supported by the dilaton field $\Phi$.
The bulk equations of motion are given by
\begin{subequations}\label{eq:bulkEOM}
\begin{align}
    \left( \frac{Zp'}{ar^{d-2}}\right)'&=0,\label{eq:eom1}\\
    -\frac{da'}{r a}-2\Phi'^2&=\frac{1}{2}\psi'^2, \label{eq:eom2}\\
    r^d a \left(\frac{(ab)'}{r^d a}\right)'-Z r^2p^{\prime 2} &= \frac{a^2}{d} (\p_i\psi)^2,\label{eq:eom3}\\
    4\left(\frac{b}{r^d}\Phi'\right)'-\frac{a}{r^{d+2}}\p_{\Phi}V + \frac{p^{\prime 2}}{2a r^{d-2}}\p_\Phi Z & = \frac{a}{2 r^{d+2}}\p_\Phi B \psi^2,\label{eq:eom4}
\end{align}
\end{subequations}
In order, they come from the $t$ components of Maxwell's equation, the $rr+tt$ and $tt+ii$ components of Einstein's equation, and the dilaton equation.  Note that we have divided by $1/d$ in the third equation above: assuming that the disorder is isotropic, any one component $\partial_x\psi \partial_x\psi \approx \frac{1}{d} \partial_i \psi \partial_i  \psi$ (with sum over $i$). Together with $\psi$'s equation of motion: 
\begin{align}\label{eq:eom scalar}
    \psi''(r)+\frac{r b'(r)-d b(r)}{rb(r)}\psi'(r)-\left(\frac{a(r)}{b(r)}k^2+ \frac{B_0}{r^2} \right)\psi(r)=0,
\end{align}
we can determine the bulk fields in the scaling limit described in the main text.

First, let us assume a renormalized metric with constant scaling exponents $\tilde z$ and $\tilde \theta$, that may differ from their UV values. The key idea is that disorder at wave number $k$ will be sensitive to the effective values of the exponents at length scale $k^{-1}$.  Since we will show (as part of our self-consistency check) that $\tilde z$ and $\tilde \theta$ vary slowly relative to $k^{-1}$ (and the bulk scalar field), we will ultimately justify this approximation.  Moving forward and solving the scalar equation of motion \eqref{eq:eom scalar} subject to the asymptotic boundary condition $\psi(r\to 0)\sim r^\# h(\vect k)$, we find
\begin{align}\label{eq:psi}
    \psi(r)\approx h(\vect k) \frac{2^{1-\tilde\sigma}\left(\frac{d-\tilde\theta}{d}\right)^{\tilde\sigma}}{\Gamma[\tilde\sigma]} r^{\frac{d(\tilde z+d-\tilde\theta)}{2(d-\tilde\theta)}}k^{\tilde\sigma}K_{\tilde\sigma}\left(\frac{d-\tilde\theta}{d}k r^{\frac{d}{d-\tilde\theta}}\right),
\end{align}
where 
\begin{align}\label{eq:tilde sigma}
    \tilde\sigma^2 = \left(\frac{\tilde z+d-\tilde\theta}{2}\right)^2+\left(\frac{d-\tilde\theta}{d}\right)^2B_0.
\end{align}
We demand $\tilde \sigma >0$ throughout the bulk in order to avoid alternate quantization. The operator dimension will also get renormalized and is fixed by
\begin{align}
    \tilde \Delta  = \frac{d+\tilde z}{2}+\tilde\sigma.
\end{align}
For a relevant operator, the no alternate quantization condition translates into $z>\theta$. As seen in the main text, this leads to a definite increase of $z$ under renormalization.
Since $B_0$ is a fixed coefficient, we can determine it using the disorder-free critical point with the scaling exponents $z$ and $\theta$, and we have 
\begin{align}
    B_0 = -\frac{d^2(d+2\nu)(d+2(z-\theta-\nu))}{4(d-\theta)^2}.
\end{align}
Substituting \eqref{eq:psi} into \eqref{eq:bulkEOM}, we can then solve for $\tilde z$ and $\tilde \theta$. As mentioned in the main text, it is enough to solve the disorder-averaged bulk equations of motion. To this end, we list useful expressions of disorder-averaged scalar fields. The full non-perturbative expressions are given by
\begin{subequations}\label{eq:dis average full}
\begin{align}
    \overline{\psi^2} &= D \frac{\sqrt{\pi} S_d 4^{ -  \tilde\sigma}}{(2\pi)^d} \left(\frac{d-\tilde\theta}{d}\right)^{-d} \frac{\Gamma[\frac{d}{2}]\Gamma[\frac{d}{2}+\tilde\sigma]\Gamma[\frac{d}{2}+2\tilde\sigma]}{\Gamma[\tilde\sigma]^2\Gamma[\frac{1+d}{2}+\tilde\sigma]} r^{\frac{d}{d-\tilde\theta}(\tilde z - \tilde\theta - 2\tilde\sigma)},\\
    \overline{k^2 \psi^2}&= D \frac{\sqrt{\pi} S_d 4^{ -  \tilde\sigma}}{(2\pi)^d} \left(\frac{d-\tilde\theta}{d}\right)^{-d-2} \frac{\Gamma[\frac{2+d}{2}]\Gamma[\frac{2+d}{2}+\tilde\sigma]\Gamma[\frac{2+d}{2}+2\tilde\sigma]}{\Gamma[\tilde\sigma]^2\Gamma[\frac{3+d}{2}+\tilde\sigma]} r^{\frac{d}{d-\tilde\theta}(\tilde z - \tilde\theta - 2\tilde\sigma - 2)},\\
    \overline{ \psi^{\prime 2}}& = D \frac{\sqrt{\pi} S_d 2^{ -1-2  \tilde\sigma}}{d(2\pi)^d} \left(\frac{d-\tilde\theta}{d}\right)^{-d-2}\left(\frac{d(2+d)(d+2\tilde\sigma)}{1+d+2\tilde\sigma} - (d+2\tilde\sigma+\tilde\theta - \tilde z)(d-2\tilde\sigma-\tilde\theta + \tilde z) \right) \nonumber\\
    &\quad \times \frac{\Gamma[1+\frac{d}{2}]\Gamma[\frac{d}{2}+\tilde\sigma]\Gamma[\frac{d}{2}+2\tilde\sigma]}{\Gamma[\tilde\sigma]^2\Gamma[\frac{1+d}{2}+\tilde\sigma]} r^{\frac{d}{d-\tilde\theta}(\tilde z - \tilde\theta - 2\tilde\sigma)-2},
\end{align}
\end{subequations}
where
\begin{equation}
    S_d = \frac{2\pi^{\frac{d}{2}}}{\Gamma(\frac{d}{2})},
\end{equation}
By keeping only linear terms in a small $\nu$ expansion, we have
\begin{subequations}\label{eq:dis average}
\begin{align}
    \overline{\psi^2} &\approx D \frac{\sqrt{\pi} S_d 2^{\theta - z}}{(2\pi)^d} \left(\frac{d-\theta}{d}\right)^{-d} \frac{\Gamma[\frac{d}{2}]\Gamma[\frac{d+z-\theta}{2}\sigma]\Gamma[\frac{d}{2}+z-\theta]}{\Gamma[\frac{z-\theta}{2}]^2\Gamma[\frac{1+d+z-\theta}{2}]} r^{\frac{2d\nu}{d-\theta}+\frac{d}{2}\gamma_a - \frac{d(z-\theta+2)}{2(z-\theta)}\gamma_b},\\
    \overline{k^2 \psi^2}&\approx D \frac{\sqrt{\pi} S_d 2^{\theta - z}}{(2\pi)^d} \left(\frac{d-\theta}{d}\right)^{-d-2} \frac{\Gamma[\frac{2+d}{2}]\Gamma[\frac{2+d+z-\theta}{2}\sigma]\Gamma[\frac{2+d}{2}+z-\theta]}{\Gamma[\frac{z-\theta}{2}]^2\Gamma[\frac{3+d+z-\theta}{2}]} r^{\frac{2d(-1+\nu)}{d-\theta}+\frac{d+2}{2}\gamma_a - \frac{(2+d)(z-\theta)+2d}{2(z-\theta)}\gamma_b},\\
    \overline{ \psi^{\prime 2}}&\approx  D \frac{d\sqrt{\pi} S_d 2^{ -2+\theta-z}}{(2\pi)^d} \left(\frac{d-\theta}{d}\right)^{-d-2}\frac{\Gamma[\frac{d}{2}]\Gamma[\frac{d+z-\theta}{2}]\Gamma[\frac{2+d}{2}+z-\theta]}{\Gamma[\frac{z-\theta}{2}]^2\Gamma[\frac{3+d+z-\theta}{2}]}  r^{\frac{2d\nu}{d-\theta}+\frac{d}{2}\gamma_a - \frac{d(z-\theta+2)}{2(z-\theta)}\gamma_b-2},
\end{align}
\end{subequations}
where we used the transformation 
\begin{subequations}\label{eq:z th}
\begin{align}
& \tilde{z}=z+\frac{d-\theta}{2 d}\left(\gamma_a z-\gamma_b(z-2)\right), \\
& \tilde{\theta}=\theta-\frac{(d-\theta)^2}{2 d}\left(\gamma_a-\gamma_b\right) .
\end{align}
\end{subequations}

At finite charge density, we find $\gamma_a\approx \gamma_b\approx\gamma_p\equiv \gamma$ in order for both \eqref{eq:eom3} and \eqref{eq:eom4} to have a consistent $r$-scaling. This is because different terms appearing in the same equation should have the same $r$-dependence, otherwise we would have nonvanishing $O(1)$ terms which is inconsistent with disorders of $O(\nu)$. \eqref{eq:eom2} results in the first line of \eqref{eq:gammaeq}
with a constant
\begin{align}\label{eq:A constant}
    A(z,\theta)= \frac{\sqrt{\pi} S_d 2^{ -3+\theta-z}}{(2\pi)^d} \left(\frac{d-\theta}{d}\right)^{-d-2}\frac{\Gamma[\frac{d}{2}]\Gamma[\frac{d+z-\theta}{2}]\Gamma[\frac{2+d}{2}+z-\theta]}{\Gamma[\frac{z-\theta}{2}]^2\Gamma[\frac{3+d+z-\theta}{2}]} .
\end{align}
We will come back to recover the whole \eqref{eq:gammaeq}.
In the main text, we give a complete solution to \eqref{eq:gammaeq} using dominant balance and extract the IR behavior from it.
Here, we find that the physics at $r\gg r_c$ can be directly inferred from a self-consistent solution of (the first line of) \eqref{eq:gammaeq}:
\begin{align}\label{eq:fp app}
    \gamma(r) &= A(z,\theta)D^* + \frac{c_1}{\log r} \equiv \gamma^*+\frac{c_1}{\log r},\nonumber\\
    \gamma^* &=  \frac{2\nu(z-\theta)}{(d-\theta)},\nonumber\\
    c_1 &= \frac{z-\theta}{d}\log \frac{D}{D^*},
\end{align}
and $\gamma^* = \gamma(r\to \infty)$ and $D^* = D_{\mathrm{eff}}(r\to\infty)$.  
To obtain the second line of \eqref{eq:gammaeq}, it is helpful to parameterize $a(r)$ as
\begin{align}\label{eq:da prefactor}
    a(r) = \left(1+\delta a(r)\right)r^{-\frac{d(z-1)-\theta}{d-\theta}-\gamma_a(r)},
\end{align}
where $\delta a(r)\sim O(\nu)$ indicates the flow of the prefactor.  Notice that we may now fix $\gamma_a(r)=\gamma(r)$ to be exactly given by (\ref{eq:gamma rel max full}), since all corrections are parameterized by $\delta a(r)$. Using the Maxwell equation \eqref{eq:eom1} to fix $\delta a(r)$ in terms of $\gamma(r)$, we find \eqref{eq:gammaeq} changes to
\begin{align}
    \gamma(r)+r\log (r) \gamma'(r) - \frac{d-\theta}{d(z+d-\theta)}r\p_r(\gamma(r)+r\log (r) \gamma'(r))= A(z,\theta) D  r^{\frac{2d\nu}{d-\theta}- \frac{d}{z-\theta}\gamma(r)}.
\end{align}
Hence, we recovered the second line of \eqref{eq:gammaeq}. To justify the dominant balance, we find the $r\p_r$ term is negligible compared to the leading term:
\begin{align}\label{eq:dominant balance}
    r\p_r(\gamma(r)+r\log (r) \gamma'(r))\sim r\p_r D_{\mathrm{eff}}(r)\sim r\p_r \left(\mathrm{const.}+r^{-\frac{2d\nu}{d-\theta}}\right)\ll D_{\mathrm{eff}}(r).
\end{align}

While \eqref{eq:gammaeq} comes from one of the four equations of motion, \eqref{eq:gamma rel max full} does not exactly solve the other equations of motion. For instance, \eqref{eq:eom3} leads to
\begin{align}
    \gamma(r)+r\log (r) \gamma'(r) - \frac{d-\theta}{d(1+d-\theta)}r\p_r(\gamma(r)+r\log (r) \gamma'(r))= B(z,\theta) D  r^{\frac{2d\nu}{d-\theta}- \frac{d}{z-\theta}\gamma(r)},
\end{align}
where 
\begin{align}
     B(z,\theta)= \frac{\sqrt{\pi} S_d 2^{-1+\theta - z}}{d(1+d-\theta)(2\pi)^d} \left(\frac{d-\theta}{d}\right)^{-d-1} \frac{\Gamma[\frac{2+d}{2}]\Gamma[\frac{2+d+z-\theta}{2}\sigma]\Gamma[\frac{2+d}{2}+z-\theta]}{\Gamma[\frac{z-\theta}{2}]^2\Gamma[\frac{3+d+z-\theta}{2}]} .
\end{align}
According to \eqref{eq:dominant balance}, one can solve it by ignoring the $r\p_r$ term and using dominant balance, but we will get a distinct $\gamma_{\mathrm{new}}(r)$ from \eqref{eq:gamma rel max full} due to $B(z,\theta)\neq A(z,\theta)$. Before solving it exactly, we can quickly see that $\gamma_{\mathrm{new}}(r)$ will not affect the IR fixed point since the difference
\begin{align}
    \gamma_{\mathrm{new}}(r) - \gamma(r) = - \frac{z-\theta}{d\log r} \log \frac{A(z,\theta)}{B(z,\theta)} \to 0,\quad r\to \infty
\end{align}
is negligible at IR.

To resolve the discrepancy and find a more accurate solution to all equations of motion  in \eqref{eq:bulkEOM}, we consider perturbative corrections to the ``constant" prefactors of $r^{-\#}$: we take \eqref{eq:da prefactor} together with
\begin{subequations}
\begin{align}
    b(r) &\approx \left(1+\delta b(r)\right)r^{-\frac{d(z-1)+\theta}{d-\theta}-\gamma_b(r)},\\
    p(r)&\approx 2\frac{d(z-1)}{d-\theta} \left( 1+\delta p(r)\right)r^{-\frac{d(z+d-\theta)}{d-\theta}-\gamma_p(r)}.
\end{align}
\end{subequations}
Since only $\delta a(r)$ enters \eqref{eq:gammaeq}, we will see that similar to the above, \eqref{eq:gamma rel max full} remains valid using the argument of dominant balance.
Substituting into \eqref{eq:eom1}, \eqref{eq:eom3} and \eqref{eq:eom4}, and using \eqref{eq:gamma rel max full}, we find
\begin{subequations}\label{eq:flow of prefactors}
    \begin{align}
        \delta a(r) = C_1 D_{\mathrm{eff}}(r),\\ 
        d(d+z-\theta)\delta b(r) - (d-\theta)r\delta b'(r) = C_2 D_{\mathrm{eff}}(r),\\
        d(d+z-\theta)\delta p(r) - (d-\theta)r\delta p'(r) = C_3 D_{\mathrm{eff}}(r),
    \end{align}
\end{subequations}
where $C_{1,2,3}$ are constants just like $A,B$ whose expressions are not illuminating, and $D_{\mathrm{eff}}(r)$ is given by \eqref{eq:Dr flow}. In the equations of motion, one generally will encounter additional terms that are precisely $r$-derivatives of the l.h.s. of \eqref{eq:flow of prefactors}, but those are negligible due to \eqref{eq:dominant balance}. Using \eqref{eq:Dstar}, the solutions in the limit $r\to \infty$ are given by
\begin{subequations}
    \begin{align}
        \delta a(r\to \infty) &= 2C_1 \frac{z-\theta}{d-\theta} \nu,\\
        \delta b(r\to \infty) &= 2C_2 \frac{z-\theta}{d(d+z-\theta)(d-\theta)} \nu,\\
        \delta p(r\to \infty) &= 2C_3 \frac{z-\theta}{d(d+z-\theta)(d-\theta)} \nu.
    \end{align}
\end{subequations}
Importantly, these prefactors will not affect the scaling exponents $z,\theta$, thus, they do not affect the IR fixed point.

Finally, the IR fixed point is controlled by the order $O(\nu)$, hence, as long as $\nu\ll 1$, it is reasonable to neglect higher-order terms $O(\nu^2)$ appearing in the equations of motion.

\section{Numerical solution of the spatially homogeneous gravitational equations}\label{app:numerics}

In this section, we perform a controlled numerical calculation of \emph{spatially homogeneous} Einstein equations (in the field theory directions, but not the bulk radial direction).  A full solution of the Einstein equations with spatial inhomogeneity is an extraordinary challenge with rather limited results (often only at fairly high temperature) \cite{Dias:2015nua}.  In this appendix, our aim is to numerically demonstrate that there is a unique finite disorder fixed point consistent with the assumption of statistical stationarity, and that it is (up to nonlinear corrections in $\nu$) identical to the one predicted in the main text.

To perform the simplest consistency check, we assume that there is Harris-relevant disorder added to a clean fixed point with $d=z=1$ and $\theta=0$.  This means that the only dynamical variables are the metric components $a(r)$ and $b(r)$, and the disordered scalar field modes $\psi(k,r)$ at each wave number. Before moving on to the details, let us summarize the numerical scheme. We use the Newton-Raphson method (see e.g. \cite{Dias:2015nua}) to solve bulk equations for $a(r)$, $b(r)$ and $\psi(k,r)$ iteratively based on disorder-free solutions.   As we have seen in the main text, the \emph{scale} $E_c$ in the bulk at which we expect to crossover to the true IR fixed point is non-perturbatively large, and is prohibitively difficult to access in numerics.   Therefore, we opt to instead look directly for the fixed point solution, by looking for self-consistent scaling solutions directly in the IR geometry. (Note that because we are directly probing the IR fixed point, the value of the radial coordinate $r$ will no longer carry much meaning, since the IR fixed point is scale-invariant!)  One difficulty becomes that a priori, we do not know any boundary conditions on any bulk fields, save for regularity in the IR.  To resolve the problem, we randomly search the UV and IR boundary conditions in our numerical integration domain that leads to the solution to the equation of motion most consistent with a scaling theory.  The simple random search converges quickly for deep and unique minimum in our calculation.   However, due to the computational expense of needing to perform these random searches for a large number of $\psi(k,r)$ fields (whose equations of motion have solutions $\sim \mathrm{e}^{\pm kr}$ and hence are numerically very sensitive to boundary conditions), we are restricted to relatively short domain sizes in bulk coordinate $r$.

Having sketched our strategy above, we observe that we now need to solve two nonlinear bulk ODEs \eqref{eq:eom2} and \eqref{eq:eom3} for $a(r)$ and $b(r)$, and one linear scalar ODE \eqref{eq:eom scalar}. Given the analysis in Appendix \ref{app:holography}, we take the following parameterization 
\begin{align}
    a(r) = \alpha(r) r^{-z(r)+1}, \quad b(r) = r^{-z(r)+1},
\end{align}
so that $\alpha(r)$ and $z(r)$ are the two functions we will solve for. Let us denote the bulk ODEs as $E_j(r; \alpha,z)=\overline{T_j(\psi)},\;j=1,2$, where $\overline{T_j}$ are averaged disorder stress tensors.  We start our numerical algorithm with $\alpha^{(0)}(r) = \hat\alpha^{(0)} = 1$, $z^{(0)}(r) = z^{(0)}=z_0$, where we use variables with hats to denote constants. Notice that even though only $z_0 = 1$ corresponds to the CFT limit, we have checked that the choice of $z_0\geq 1$ in numerics does not change the eventual fixed point of the algorithm. The domain for ODE is denoted as $[r_i,r_f]$. We now proceed as follows:
\begin{enumerate}
    \item \label{it:1} Solve the scalar ODE with $\alpha^{(n)}(r), z^{(n)}(r) $ subject to the initial condition $[\psi_{z^{(n)}},\; \p_r\psi_{z^{(n)}}](k,r_i)$, where $r_i$ is the initial point and $\psi_{z^{(n)}}$ is determined through \eqref{eq:psi} with $\tilde z = z^{(n)}$. For each wavevector $k$, we obtain $\psi^{(n)}(k,r)$. The disorder average is performed by summing over different $k$, and the disorder strength is fixed to be $D=D_0$.
    \item \label{it:2} Now we solve the bulk equations of motion. 
 Randomly search initial conditions by 
    \begin{align}\label{eq:ini condition}
        [\delta z,\delta z',\delta \alpha](r_i) = \left\{ [\delta \hat z,0,\delta \hat \alpha]\in \mathcal{B}|\min \left( E_j(\hat \alpha^{(n)}+\delta \hat{\alpha},\hat z^{(n)}+\delta \hat{z}) - \overline{T_j(\psi^{(n)})}\right)\right\},
    \end{align}
    where $\mathcal{B}$ is a domain from which our initial conditions are drawn uniformly.  
    \item \label{it:3} Expand the bulk ODEs by $\alpha = \alpha^{(n)}+\delta \alpha, \; z = z^{(n)}+\delta z$, and
    \begin{align}
        E_j(\alpha^{(n)},z^{(n)}) + \frac{\delta E_j}{\delta \alpha}(\alpha^{(n)},z^{(n)}) \delta \alpha(r) + \frac{\delta E_j}{\delta z}(\alpha^{(n)},z^{(n)}) \delta z(r)  = \overline{T_j(\psi^{(n)})},
    \end{align}
    where $\delta E_j/\delta \alpha, \delta E_j/\delta z$ are differential operators \cite{Dias:2015nua}. We solve for $\delta \alpha,\delta z$ subject to the initial condition \eqref{eq:ini condition}.
    \item \label{it:4} Update various variables following
    \begin{align}
        \alpha^{(n+1)}(r) & = \alpha^{(n)}(r) +\delta \alpha(r),\nonumber\\
        z^{(n+1)}(r) & = z^{(n)}(r) +\delta z(r),\nonumber\\
        \hat \alpha^{(n+1)} & = \hat \alpha^{(n)} +\delta \alpha(r_i),\nonumber\\
        \hat z^{(n+1)} & = \hat z^{(n)} +\delta z(r_i).
    \end{align}
    Then, repeat Step \ref{it:1}, \ref{it:2}, \ref{it:3} with the new variables until the convergence is reached.
\end{enumerate}
There are two sets of variables under iteration, the hatted $\hat\alpha,\hat z$ and un-hatted $\alpha(r),z(r)$. We emphasize that the hatted ones are constant in $r$, while the un-hatted ones are solutions to the linearized ODEs. The reason to use hatted variables in random search is that we wish both $\alpha(r),z(r)$ to reach a fixed point so that they will not depend on $r$, and the target function in \eqref{eq:ini condition} is exactly to achieve the goal. At the same time, we fix the initial condition for the first-order derivative to be zero $\delta z' (r_i)= 0$ in order to be consistent with the fixed point solution.

\begin{figure}[t]
\includegraphics[width=.95 \linewidth]{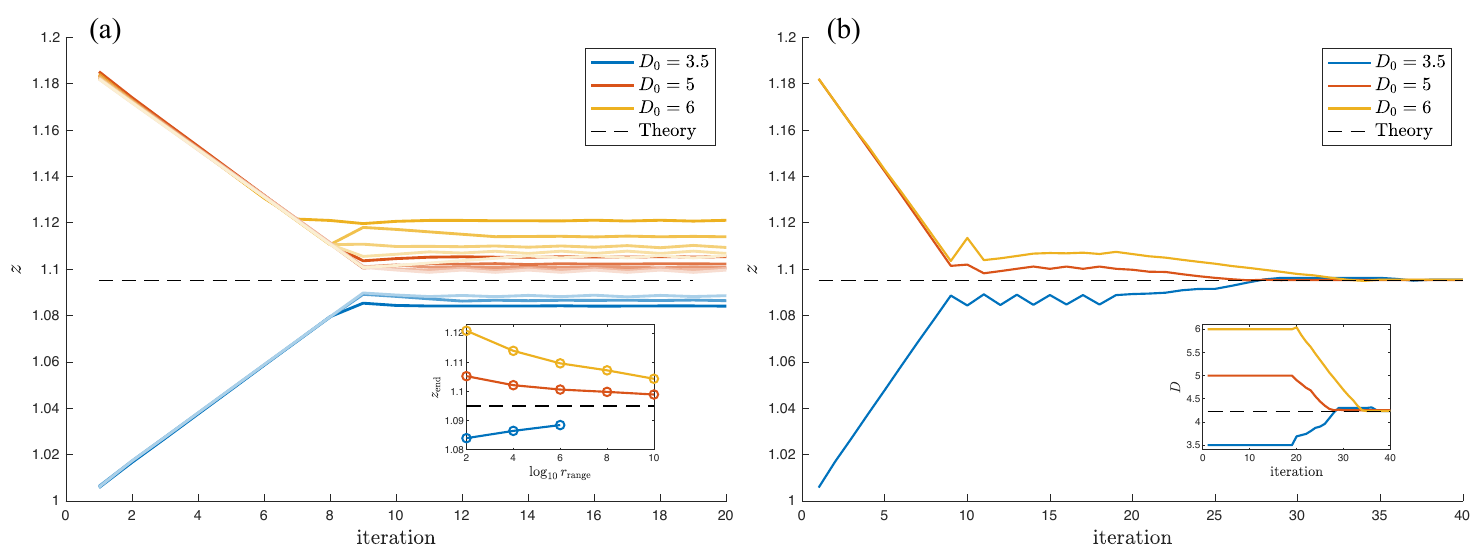}
\caption{Numerical results for constructing the disordered fixed point geometry, based on iterative Newton-Raphson method. We choose $\nu=0.05$ and $\mathcal{B} = [-0.01,0.01]$ with 5000 random searches. (a) Evolution of the mean value of $z(r)$ based on Step \ref{it:1} to \ref{it:4}. Different initial conditions are chosen with $D_0=3.5,z_0=1$ (blue), $D_0=5, z_0=1.2$ (red), and $D_0=6, z_0=1.2$ (yellow). From dark to light colors, they correspond to different ranges of $r$: $[1,2]\times 10^2 - 10^{10}$; see inset for the end point of $z$ against different ranges of $r$. Black dashed line indicates the nonlinear fixed point predicted by our theory \eqref{eq:full z}. (b) evolution of the mean value of $z(r)$ with updated disorder strength $D$. 
Updates of $D$ begin at iteration 20 with $\mathcal{B}_D = [-0.2,0.2]$. For $D_0=3.5$, we needed to reject updates that drive the system towards the disorder free fixed point $D=0$, $z=1$ for the first such iterations.  The black dashed line in the inset denotes the analytic fixed point in \eqref{eq:full D}.}
\label{fig:iteration}
\end{figure}

We see that both $\alpha(r),z(r)$ will approach a constant value once convergence is achieved; see Figure \ref{fig:iteration}(a) for the evolution of the mean value of $z(r)$. To compare to our theory, we use the non-perturbative disorder average in \eqref{eq:dis average full}. Following the analysis in Appendix \ref{app:holography},  we find the full \emph{nonlinear} fixed point satisfies
\begin{align}
    z^* - 2\sigma^* = 0.
\end{align}
In Figure \ref{fig:iteration}, we chose $\nu=0.05$, such that 
\begin{align}\label{eq:full z}
    z^*_{\mathrm{full}} = 1.095.
\end{align}
Recall that the linear fixed point is $z^*_{\mathrm{linear}} = 1 + 2\nu = 1.1$.  We plot the convergence of $z(r)$ for different ranges of $r$ domain. The inset of Fig.\ref{fig:iteration}(a) shows that by increasing $r_{\mathrm{range}}$ the end point of $z(r)$ will approach some fixed values at a slow logarithmic rate. The reason is that the algorithm indeed converges to some approximate fixed point, but if the value of $D=D_0$ is held fixed, it is not able to find the true fixed point -- thus, the precise domain of $r$ in which we solve for modifies the ultimate value of $z$.   

To find a true scale invariant fixed point, at which the scale $r$ drops out of the final critical exponents, we must further allow the value of $D$ to change in the iterations. We thus perform additional steps as follows. Choosing $D^{(0)} = D_0$,
\begin{itemize}
    \item [$2'$.] \label{it:2new} In addition to Step \ref{it:2}, we perturb the disorder strength by $D^{(n)} + \delta D$ and optimize over $\delta D$ for each realization of $\mathcal{B}$. Specifically, the initial condition becomes
    \begin{align}\label{eq:ini condition 2}
        [\delta z,\delta z',\delta \alpha](r_i) = \left\{ [\delta \hat z,0,\delta \hat \alpha]\in \mathcal{B}|\min \left( E_j(\hat \alpha^{(n)}+\delta \hat{\alpha},\hat z^{(n)}+\delta \hat{z}) - \overline{T_j(\psi^{(n)})}_{D^{(n)}+\delta D_{(\delta \hat{\alpha},\delta \hat z)}}\right)\right\},
    \end{align}
    where
    \begin{align}
        \delta D_{(\delta \hat{\alpha},\delta \hat z)} = \left\{ \delta D\in \mathcal{B}_D|\min \left( E_j(\hat \alpha^{(n)}+\delta \hat{\alpha},\hat z^{(n)}+\delta \hat{z}) - \overline{T_j(\psi^{(n)})}_{D^{(n)}+\delta D}\right)\right\}.
    \end{align}

    \item [$4'$.] \label{it:4new} We only update the fields if the minimum value of the random search in \eqref{eq:ini condition 2} is smaller than the one in the previous iteration.
    The disorder strength is updated through 
    \begin{align}
        D^{(n+1)} = D^{(n)} + \delta D_{(\delta \hat{\alpha} = \delta \alpha(r_i),\delta \hat z = \delta z(r_i))}.
    \end{align}
\end{itemize}
Based on Figure \ref{fig:iteration}(a), we start to update $D$ after iteration 20 and obtain Figure \ref{fig:iteration}(b). We see that all the curves converge to the fixed point predicted by our theory \eqref{eq:full z}, and, at the same time, the inset shows that the disorder strength will also converge to the same value. This demonstrates that there is indeed a unique stable fixed point, consistent with homogeneous geometry.  

Now let us compare this fixed point to our theoretical predictions. The prediction in \eqref{eq:Dstar} is valid to the first order in $\nu$. The full non-perturbative result is (assuming $z>1,\theta\neq 0$ at the clean fixed point)
\begin{align}
    D^* = \frac{d(z^*-1)-\theta}{A(z^*,\theta)(d-\theta)}.
\end{align}
This leads to the fixed point of disorder strength in the CFT limit as
\begin{align}\label{eq:full D}
    D^*_{\mathrm{full}} = \frac{z^*_{\mathrm{full}} - 1}{A(z^*_{\mathrm{full}},\theta = 0)} \approx 4.23,
\end{align}
where we used \eqref{eq:full z} and \eqref{eq:A constant}. We have confirmed that the end point of $D$ shown in the inset of Figure \ref{fig:iteration}(b) agrees perfectly with \eqref{eq:full D} .

\section{Charge-neutral disordered fixed point}\label{app:charge neutral}
In this appendix, we discuss how the calculation is modified for charge-neutral systems.  Since the steps are analogous to what we did for charged black holes, we will be relatively brief.

At charge neutrality by turning off the Maxwell field, we have the following equations of motion
\begin{subequations}
    \begin{align}
        2\gamma_a(r)-\gamma_b(r)+r\log (r) (2\gamma_a'(r)-\gamma_b'(r))&= A(z=1,\theta) D  r^{\frac{2d\nu}{d-\theta}+\frac{d}{2}\gamma_a - \frac{d(3-\theta)}{2(1-\theta)}\gamma_b},\\
        \gamma_a(r)+\gamma_b(r)+r\log (r) (\gamma_a'(r)+\gamma_b'(r))& \nonumber\\
        - \frac{d-\theta}{d(1+d-\theta)}r\p_r(\gamma_a(r)+\gamma_b(r)+r\log (r) (\gamma_a'(r)+\gamma_b'(r)))&= 2B(z=1,\theta) D  r^{\frac{2d\nu}{d-\theta}+\frac{d}{2}\gamma_a - \frac{d(3-\theta)}{2(1-\theta)}\gamma_b},
    \end{align}
\end{subequations}
where, in order, they come from \eqref{eq:eom2} and \eqref{eq:eom3}. In the above equation, we used $c_\Phi(r) \equiv c_{\Phi,0} - \gamma_\Phi$, where $c_{\Phi,0}$ is given in \eqref{eq:scaling form},  and $\beta\gamma_{\Phi} = \gamma_a-\gamma_b$; the latter constraint comes from \eqref{eq:mass term}. 
 Here, we present the self-consistent solution as the IR fixed point, and argue for its stability via the same dominant balance arguments as we described above.  We find 
\begin{align}
    \gamma_a(r)+\gamma_b(r)& = 2 B(z=1,\theta)D^*+ \frac{2c_1}{\log r}  \equiv \gamma_a^*+\gamma_b^* + \frac{2c_1}{\log r},\nonumber\\
    2\gamma_a(r)-\gamma_b(r)& = A(z=1,\theta) D^* + \frac{c_1}{\log r} \equiv 2\gamma_a^*-\gamma_b^* + \frac{c_1}{\log r},\nonumber\\
    0&=\frac{2d\nu}{d-\theta}+\frac{d}{2}\gamma^*_a - \frac{d(3-\theta)}{2(1-\theta)}\gamma_b^*,\nonumber\\
    c_1&= \frac{1-\theta}{d}\log \frac{D}{D^*}.
\end{align}
Note, we assume that both $\gamma_a(r)$ and $\gamma_b(r)$ will approach their IR fixed point at the same rate. This is because their difference $\propto 1/\log r$ will only change the cut-off through the constraint $\gamma_\Phi\propto \gamma_a-\gamma_b$ so is not important in the scaling limit.
Solving the above, we obtain 
\begin{subequations}
\begin{align}
     \gamma_a^* & = \frac{2 \nu (1-\theta) (3 d-2 \theta)}{(d-\theta) (3 d+(\theta-5) \theta)},\\
     \gamma_b^* &= \frac{2 \nu (1-\theta) (3 d-4 \theta)}{(d-\theta) (3 d+(\theta-5) \theta)},\\
     \gamma_\Phi^* &= \frac{4 \nu (1-\theta) \theta}{(d-\theta) (3 d+(\theta-5) \theta)} \sqrt{\frac{-d(d-\theta)}{8\theta}}.
\end{align}
\end{subequations}
Note that $\theta<0$ due to \eqref{eq:null energy condition}.
Using \eqref{eq:z th}, we arrive at the IR geometry \eqref{eq:FPrel no maxwell}.

\section{Details of the holographic calculation of conductivity}\label{app:conductivity}
This appendix contains details of the holographic calculation of conductivity.  Here we assume that the system is at finite density, and that $z^*<\infty$.  Thermoelectric conductivities were calculated in rather general inhomogeneous backgrounds in \cite{Banks:2015wha}.  As we explained in the main text, the dominant inhomogeneity is only in the bulk scalar $\psi$, and so at leading order at small $\nu$, the dc thermoelectric conductivity matrix is given by
\begin{subequations}\label{eq:thermoelectric}
\begin{align}
    \sigma_{\mathrm{dc}}&= \frac{s Z^{(0)}r_+^2}{4\pi }+\frac{4\pi \rho^2}{s \langle (\p_i\psi(\vect x))^2\rangle },\\
    \alpha_{\mathrm{dc}} &= \frac{4\pi \rho}{ \langle (\p_i\psi(\vect x))^2\rangle },\\
    \bar\kappa_{\mathrm{dc}} &= \frac{4\pi T s}{ \langle (\p_i\psi(\vect x))^2\rangle },
\end{align}
\end{subequations}
where \begin{equation}
    \langle (\p_i\psi(\vect x))^2\rangle = \frac{1}{V}\int \mathrm{d}^dx \left(\partial_i \psi(\vect x, r_+)\right)^2.
\end{equation}
where $V$ is the spatial volume in the boundary theory.
The terms inversely proportional to $\langle (\p_i\psi(\vect x))^2\rangle$ will dominate, and it is useful to denote the relaxation rate as
\begin{align}
    \Gamma =  \frac{s}{4\pi} \langle (\p_i\psi(\vect x))^2\rangle.
\end{align}
Various $T$-scalings are shown in \eqref{eq:dc drude}. 

There are also contributions to the conductivity that do not depend explicitly on the disorder (beyond how disorder flows to a particular fixed point with fixed $z^*$ and $\theta^*$!).
Applying \eqref{eq:thermoelectric} to calculate the incoherent conductivity that is insensitive to momentum relaxation, we find
\begin{align}
    \sigma_{\mathrm{dc,inc}} =  \frac{(sT)^2 \sigma_{\mathrm{dc}} - 2sT\rho \alpha_{\mathrm{dc}}+\rho^2 T\bar\kappa_{\mathrm{dc}}}{\mathcal{M}^2} =  \left(\frac{sT}{\mathcal{M}}\right)^2 \frac{s Z^{(0)}r_+^2}{4\pi } \sim T^{2+\frac{d-\theta^*-2}{z^*}},
\end{align}
since, as we will argue later, for generic models we expect the constant prefactor $\mathcal{M}\sim T^0$.

To obtain the optical conductivity, we need to solve the perturbed bulk equations of motion.
Consider perturbing the system by a small AC electric field along the $\hat{x}$ direction. Such a perturbation couples to
\begin{subequations}\label{eq:delta Ax}
    \begin{align}
    \delta g_{tx} &= \frac{h_{tx}(r)}{r^2}\e^{-\ii\omega t}, \\
     \delta A &= \delta A_x(r)\e^{-\ii\omega t} \ud x, \\
     \delta\psi &= \psi(r,\vect{x})\delta P(r,\vect x) \e^{-\ii\omega t}, 
\end{align}   
\end{subequations}
where $\psi(r,\vect{x})$ is the background inhomogeneous scalar field. The $rx$-component of Einstein’s equations, the $x$-component of Maxwell's equation, and the scalar equation read 
\begin{subequations}\label{eq:optical bulk ode}
\begin{align}
    \frac{h_{tx}'}{r^d a}-\rho \delta A_x + \frac{b }{\omega r^d} \int \frac{\ud^d k}{(2\pi)^d} k_x \psi^2(\vect k) \delta P'(\vect k)&=0,\label{eq:rx optical}\\
    \left(-\rho h_{tx}+r^{2-d}Z b \delta A_x'\right)'+\frac{r^{2-d}Z \omega^2}{b}\delta A_x &=0,\label{eq:x optical}\\
    \left(\frac{b \psi^2(\vect k)}{r^d}\delta P'(\vect k)\right)' + \frac{ \omega k_x }{r^d  b}\psi^2(\vect k) h_{tx}+\frac{\omega^2}{r^{d} b}\psi^2(\vect k)\delta P(\vect k)&=0, \label{eq:scalar optical}
\end{align}
\end{subequations}
where $\delta A_x=\delta A_x(\vect k=0)$ and $h_{tx}=h_{tx}(\vect k=0)$ are at zero momentum. In deriving \eqref{eq:rx optical}, we ignored the term $\p_x\psi\psi'\delta P$, which can be regarded as a higher order correction. 

At high frequency, $\omega\gg T$, the term proportional to $\delta P$ in \eqref{eq:rx optical} is perturbatively small for weak disorder $\psi^2\propto D^*\ll 1$. Meanwhile, this term has the same $r$-dependence, $r^{-\frac{d^2-d\theta^*+d}{d-\theta^*}}$, as the other terms in \eqref{eq:rx optical} since the disorder is exactly marginal at the IR fixed point. Hence, we can approximate $\delta P=0$ at leading order, and the resulting system of ODEs is closed for $\delta A_x$ and $h_{tx}$. Combining \eqref{eq:rx optical} and \eqref{eq:x optical}, we have
\begin{align}
    \left(r^{2-d}Z b \delta A_x'\right)' + \left( \frac{r^{2-d}Z}{b}\omega^2 - r^d a \rho^2\right) \delta A_x = 0.
\end{align}
Applying the change of variables,
\begin{align}
    \frac{\ud w}{\ud r}=\frac{1}{b},\quad  \delta \bar A_x = \sqrt{r^{2-d}Z}\delta A_x,
\end{align}
we obtain
\begin{align}
    \left(\p_w^2+\omega^2\right)\delta\bar A_x + c_{1} \delta \bar A_x =0,
\end{align}
where
\begin{align}
    c_{1}=- \frac{r^d a b \rho^2}{r^{2-d}Z} - \frac{b}{\sqrt{r^{2-d}Z}}\left(\frac{b(r^{2-d}Z)'}{2\sqrt{r^{2-d}Z}}\right)'.
\end{align}
We find $c_1 \sim w^{-2}$ meaning the solution can be written as $\delta \bar A_x = G(w\omega)$. Applying the holographic dictionary, together with a matching argument \cite{hartnoll_2018_holographic} to connect with the UV scaling, we obtain
\begin{align}
    \re \sigma(\omega\gg T) = \frac{1}{\omega} \im G^R_{\delta  A_x \delta  A_x} \sim \omega^{2+\frac{d-\theta^*-2}{z^*}}.
\end{align}

At low frequency $\omega\ll T$, however, we cannot neglect the contribution of $\delta P$ in \eqref{eq:rx optical}, and it will contribute to the relaxation time $\tau \propto D^{*-1}$ in the ac conductivity.  Following \cite{Lucas:2015vna}, we work in the limit $\omega/T\to 0$, but keep $\omega \nu^{-1}$ finite. Define
\begin{align}
    \delta \mathcal{P}_x \equiv  \frac{b }{\omega r^d} \int \frac{\ud^d k}{(2\pi)^d} k_x \psi^2(\vect k) \delta P'(\vect k),
\end{align}
With this overall factor of $\omega^{-1}$, we will be able to safely take the $\omega \rightarrow 0$ limit below. 
At leading order in $\omega$, \eqref{eq:optical bulk ode} becomes
\begin{subequations}\label{eq:e12}
    \begin{align}
        \frac{1}{a r^d} h_{tx}' - \rho \delta A_x + \delta\mathcal{P}_x &=0,\\
        \rho h_{tx}' - \left( b r^{2-d}Z \delta A_x' \right)' &=0,\\
        \delta \mathcal{P}_x' + h_{tx}\left[ \frac{1}{b r^d} \int \frac{\ud^d k}{(2\pi)^d}  k_x^2 \psi(\vect k)^2\right] &=0.\label{eq:dp h eom}
    \end{align}
\end{subequations}

Let us now solve these equations subject to appropriate boundary conditions.  It is helpful to first identify all solutions without regards to boundary conditions, and then stitch together the correct solution (compatible with boundary conditions) at the end.  The first solution is given by
\begin{align}\label{eq:first mode}
    \delta A_x = \delta A_x^0,\quad \delta \mathcal{P}_x = \rho \delta A_x^0,\quad h_{tx}=0,
\end{align}
which directly couples to the disorder. The second solution of interest is the ``Galilean boost'' mode: 
\begin{align}\label{eq:boost mode}
    \delta A_x = c_1 (p(r)+p(r_+)),\quad h_{tx}=c_2 - a(r)b(r)c_1, \quad \delta \mathcal{P}_x = 0,
\end{align}
where we have modified its form in \cite{Lucas:2015vna} by noting that the coefficients $c_{1,2}$ must (from our UV theory's perspective) have differing dimensions.  It was shown in \cite{Lucas:2015vna} that the remaining two solutions to (\ref{eq:e12}) do not contribute at leading order to $\sigma(\omega)$, and the conclusion is unchanged here. Note that when the system is Lorentz invariant ($z=1$), there is no need to include $c_{1,2}$. One way to fix these parameters is to use a UV-completion of our scaling theory to AdS, in which case we would find that $c_{1,2}$ are related by the UV scale at which we crossover to an AdS UV-completion.  In this particular UV-completion, $c_{1,2}$ can be chosen to scale independently of $T$.  We expect that this conclusion is more general, although a detailed analysis requires a careful holographic renormalization calculation which is non-trivial for these Lifshitz and hyperscaling-violating backgrounds \cite{Chemissany:2014xsa,Taylor:2015glc}. 
Using $b(r\to r_+)\approx 4\pi T (r_+-r)$ and \eqref{eq:eom3}, we can determine $p(r_+) = sT/\rho$ up to $O(D)$ corrections. Now, let us start from \eqref{eq:first mode} with infalling boundary condition:
\begin{subequations}
    \begin{align}
        \delta A_x(r\to r_+)&=\delta A_x^0 \left(1+\frac{\ii \omega}{4\pi T} \log \frac{r_+}{r_+-r}\right),\\
        \delta \mathcal{P}_x(r\to r_+)&=\rho \delta A_x^0 \left(1+\frac{\ii \omega}{4\pi T} \log \frac{r_+}{r_+-r}\right).
    \end{align}
\end{subequations}
Plugging it in \eqref{eq:dp h eom} at $r\to r_+$, we find
\begin{align}
    \frac{\ii \omega}{4\pi T} \rho \delta A_x^0 \frac{1}{r_+-r} = - \frac{h_{tx}}{4\pi T (r_+-r) r_+^d} \left[  \int \frac{\ud^d k}{(2\pi)^d}  k_x^2 \psi(\vect k)^2\right].
\end{align}
Therefore, using \eqref{eq:boost mode}, the leading contributions to $\delta A_x$ are given by
\begin{align}
    \delta A_x = \delta A_x^0 (1-\mathrm{i}\omega C (p+p(r_+)),
\end{align}
where
\begin{align}
    C \equiv  \rho \left(\frac{1}{r_+^d} \int \frac{\ud^d k}{(2\pi)^d}  k_x^2 \psi(\vect k)^2\right)^{-1} \frac{c_1}{c_2}.
\end{align}
We can then define the relaxation time $\tau$ using $\delta A_x$ at UV:
\begin{align}
    \tau \equiv C (p(r=0)+p(r_+)) = \mathcal{M}\Gamma^{-1}
\end{align}
Recalling that the conductivity is determined by $\sigma(\omega)\sim \frac{\delta A^{(1)}_x}{\delta A^{(0)}_x}$, where $\delta A^{(0)}_x$ ($\delta A^{(1)}_x$) is the coefficient for the leading (subleading) order in the asymptotic expansion at $r\to 0$, we obtain the Drude peak $\sigma(\omega)\sim (1-\ii\omega \tau)^{-1}$. As long as $c_1,c_2,\mu$ are $T$-independent, $\tau$ and $\Gamma$ have the $T$-dependent scaling
\begin{align}
    \tau \sim \Gamma^{-1} \sim T^{-\frac{2+d-\theta^*}{z^*}}.
\end{align}

As explained in the main text, we can only trust the existence of the Drude peak when $\tau\gg T^{-1}$, which means $2+d-\theta^* - z^*>0$.  If this criterion does not hold, our evaluation of the IR conductivity at the horizon is exact, but we cannot controllably calculate the leading order $\omega$-dependent corrections to $\sigma(\omega)$; we would then have analytic control only at $\omega=0$ and $\omega \gg T$.

\section{Local criticality with $z=\infty$ and $\theta=-\eta z$}\label{app:local critical}

In this appendix, we generalize our calculation to the case where
\begin{align}\label{eq:z th infty}
    z\rightarrow \infty,\quad \theta = -\eta z \rightarrow -\infty,
\end{align}
with a finite and fixed $\eta>0$. The metric will then scale as
\begin{align}
    a(r)\sim r^{-\frac{d+\eta}{\eta}},\quad b(r)\sim r^{ -\frac{d-\eta}{\eta}  }.
\end{align}
As before, we consider the renormalized metric with $\eta\to \tilde\eta$. The scalar equation of motion becomes
\begin{align}
    \psi'' - \frac{d+d\tilde\eta - \tilde\eta}{\tilde\eta r}\psi' - \frac{B_0+k^2}{r^2}\psi = 0.
\end{align}
This equation admits power-law solutions
\begin{align}\label{eq:psi z th infty}
    \psi(\vect k,r) = h(\vect k) r^{\frac{d+d\tilde\eta}{2\tilde\eta}-\tilde\sigma} ,\quad \tilde\sigma^2 = \left(\frac{d+d\tilde\eta}{2\tilde\eta}\right)^2+k^2+B_0,
\end{align}
where we neglected the normalizable solution. Observe that $\omega\sim k^z$ and $r^{-1}\sim k^{(d-\theta)/d}$, therefore, under \eqref{eq:z th infty}, $k$ does not scale and the theory realizes ``local criticality'' \cite{Hartnoll:2012wm}. Based on this, we can deduce that when $B_0=0$ the disorder is Harris-marginal since $\psi(\vect k=0,r)$ has no $r$-dependence, and when $B_0<0$ ($B_0>0$), it corresponds to Harris-relevant(irrelevant) disorder. Evaluating the disorder average for a relevant disorder with $|B_0|\ll 1$, we obtain for $r\Lambda\gg 1$
\begin{align}
    \overline{\psi^2} = \int \frac{\ud^d k}{(2\pi)^d} \psi(\vect k)^2 \approx \frac{S_d}{(2\pi)^d} D \int \ud k k^{d-1} r^{-\frac{(k^2+B_0)\tilde\eta}{d+d\tilde\eta}} \sim   \frac{D}{(\log r)^{d/2}} r^{-\frac{B_0\tilde\eta}{d+d\tilde\eta}}.
\end{align}
We find that no matter how the renormalized exponent $\tilde \eta$ behaves, this contribution to the stress tensor will always blow up for relevant disorders in the deep IR, making the bulk equations of motion inconsistent. However, we do not exclude the existence of a valid IR fixed point, possibly extending our scheme to a more general renormalized geometry.

Nevertheless, it is still sensible to discuss the marginal disorder and its corresponding fixed point. When $B_0=0$, we obtain, using dominant balance,
\begin{align}\label{eq:ode z th infty}
    \gamma(r) + r\log r \gamma'(r) \sim  \frac{D}{(\log r)^{d/2}},
\end{align}
where we used the same parametrization $\gamma_a\approx \gamma_b \approx\gamma_p\equiv\gamma$ as before in Appendix \ref{app:holography}. The solution to \eqref{eq:ode z th infty} scales as
\begin{align}\label{eq:gamma local critical marginal}
    \gamma(r)\sim 
    \begin{cases}
        D \frac{2}{2-d} \frac{1}{(\log r)^{d/2}} ,\quad d\neq 2\\
        D \frac{\log(\log r)}{\log r} , \quad d=2
    \end{cases}
\end{align}
Since $\gamma(r\to \infty)\to 0$, the disorder will become marginally irrelevant at deep IR. This is similar to the conclusion made in \cite{Ganesan:2020wzm}, but the scaling \eqref{eq:gamma local critical marginal} is dramatically different. 

The marginal disorder will support to a Drude-like dc conductivity. Introducing the horizon as before, we have $T\sim r_+^{-d/\eta}$ and $s\sim T^{\eta}$. We can then evaluate the relaxation rate as
\begin{align}
    \Gamma = \frac{1}{r_+^d} \int \frac{\ud^d k}{(2\pi)^d}  k_x^2 \psi(\vect k)^2 \approx \frac{S_d}{d(2\pi)^d} \frac{D}{r_+^d} \int \ud k k^{d+1} r_+^{-\frac{k^2 \eta}{d+d\eta}}\sim  \frac{D}{r_+^d} \frac{1}{(\log r_+)^{d/2+1}}
\end{align}
Hence, for $D\ll 1$, the $T$-scaling of the dc electrical conductivity is dominated by the Drude form
\begin{align}\label{eq:sigma dc local critical}
    \sigma_{\mathrm{dc}} \sim \Gamma^{-1}\sim \frac{1}{D} T^{-\eta} (\log T )^{d/2+1}.
\end{align}
The Drude form manifests the fact that the disorder is the leading contribution to the momentum relxation, thus, even it is marginally irrelevant at the IR fixed point, the disorder is a dangerously irrelevant operator for the transport properties. 
Interestingly, when $\eta = 1$, \eqref{eq:sigma dc local critical} has a linear-in-$T$ resistivity (up to logarithm), and, at the same time, has an entropy $s\sim T$. This fixed point corresponds to the Gubser-Rocha model \cite{Gubser-Rocha}, which has been generalized to inhomogeneous charge density at fixed wave number, where numerics suggest robust linear-in-T resistivity \cite{Balm:2022bju}. 

\end{appendix}

\bibliography{EMD}

\end{document}